\documentclass[journal=jacsat,manuscript=suppinfo,layout=traditional]{achemso}

\author{Natalia A. Denesyuk}
\author{D. Thirumalai}
\email{thirum@umd.edu}
\affiliation{Department of Chemistry and Biochemistry and Biophysics Program, Institute for Physical Science and Technology,
University of Maryland, College Park, Maryland 20742}
\title{Crowding Promotes the Switch from Hairpin to Pseudoknot Conformation in Human Telomerase RNA}

\renewcommand{\baselinestretch}{2}

\begin{document}

\section {Three interaction site (TIS) model of RNA}

We develop a realistic force field for nucleic acids using the TIS model~\cite{Hyeon}, in which each nucleotide is replaced
by three spherical beads P, S and B, representing respectively a phosphate, a sugar and a base (Figure S1). The 
coarse-grained beads are at the center of mass of the chemical groups and have a radius $R_i$ of 2 \r{A} for phosphates, 
2.9 \r{A} for sugars, 2.8 \r{A} for adenines, 3 \r{A} for guanines and 2.7 \r{A} for cytosines and uracils. The values of 
$R_i$ are calculated using $V_i = 4\pi R_i^3/3$, where $V_i$ is the van der Waals volume of the chemical group computed 
from the coordinates and radii of its individual atoms. We use the total molecular weight of each RNA group as the mass of 
the representative bead in our simulations. In the TIS representation of nucleic acids, bond lengths, 
$\rho$, and valence angles, $\alpha$, are constrained by harmonic potentials, $U(\rho)=k_{\rho}(\rho-\rho_0)^2$ and 
$U(\alpha)=k_{\alpha}(\alpha-\alpha_0)^2$, where the equilibrium values $\rho_0$ and $\alpha_0$ are obtained by 
coarse-graining an ideal A-form RNA helix\cite{IdealRNAsite}. The values of $k_{\rho}$, in kcal$\,$mol$^{-1}$\r{A}$^{-2}$, 
are: 64 for an S(5')$-$P bond, 23 for an P$-$S(3') bond and 10 for an S$-$B bond. The values of $k_{\alpha}$ are 
5 kcal$\,$mol$^{-1}$rad$^{-2}$ if the valence angle involves a base, and 20 kcal$\,$mol$^{-1}$rad$^{-2}$ otherwise. 
We have chosen $k_\rho$ and $k_\alpha$ so that the time averages of $(\rho-\rho_0)^2$ and $(\alpha-\alpha_0)^2$ measured 
in simulations at $15\ ^{\circ}$C, match the corresponding quantities averaged over all bonds in the coarse-grained NMR 
structure of the hTR pseudoknot (PDB code 2K96).

Single strand stacking interactions, $U_{\mathrm {ST}}$, are applied to all pairs of consecutive nucleotides along the 
chain,
\begin{equation}
U_{\mathrm {ST}}=\frac{U_{\mathrm {ST}}^0}{1+1.4(r-r_0)^2+4(\phi_1-\phi_{10})^2+4(\phi_2-\phi_{20})^2},
\label{UST}
\end{equation}
where $r$, $\phi_1$ and $\phi_2$ are defined in Figure S1a. The equilibrium values $r_0$, $\phi_{10}$ and 
$\phi_{20}$ are extracted from the coarse-grained structure of an ideal A-form RNA helix\cite{IdealRNAsite} and depend on 
the chemical identities of the two nucleotides. We obtain the constants $U_{\mathrm{ST}}^0$ for sixteen distinct nucleotide 
dimers from available experimental data on stacking of nucleic acid bases in single-stranded and double-stranded 
RNA~\cite{DoubleStrandData,NAbook,Dima}, as described next.

Additive contributions $\Delta G{b-c\choose a-d}$ of individual stacks to the total stability of an RNA double helix, 
where $a-d$ and $b-c$ are stacked Watson-Crick base pairs, are known experimentally~\cite{DoubleStrandData}. 
For reference, experimentally determined enthalpic $\Delta H{b-c\choose a-d}$ and entropic $\Delta S{b-c\choose a-d}$ 
contributions to $\Delta G{b-c\choose a-d}$ are reproduced in Table S1. We make the following approximations:
\begin{eqnarray}
\Delta H{b-c\choose a-d}&=&\Delta H{b\choose a} + \Delta H{d\choose c} + 0.5\Delta H(a-d) + 0.5\Delta H(b-c), \nonumber\\
\Delta S{b-c\choose a-d}&=&\Delta S{b\choose a} + \Delta S{d\choose c}, 
\label{dG2}
\end{eqnarray}
where $\Delta H{b\choose a}$ and $\Delta S{b\choose a}$ are the enthalpy and entropy changes resulting from stacking of 
$b$ over $a$ along $5'\rightarrow3'$ in one strand, respectively, and $\Delta H(a-d)$ is the additional stability due to 
hydrogen bonding between $a$ and $d$ in two complementary strands. Inspection of $\Delta H{b-c\choose a-d}$ and 
$\Delta S{b-c\choose a-d}$ in Table S1 leads us to conclude that, except for ${\mathrm{C}\choose\mathrm{G}}$ and 
${\mathrm{G}\choose\mathrm{C}}$, stacking parameters do not depend strongly on the order of nucleotides along 
$5'\rightarrow3'$. Therefore, we assume, with the exception of ${\mathrm{C}\choose\mathrm{G}}$ and 
${\mathrm{G}\choose\mathrm{C}}$, that $\Delta H{b\choose a}=\Delta H{a\choose b}$ and 
$\Delta S{b\choose a}=\Delta S{a\choose b}$, which is also valid within the range of experimental uncertainties specified 
in ref 2. This assumption allows us to average the experimental values for 
$\Delta H{\mathrm{U-A}\choose\mathrm{A-U}}$ and $\Delta H{\mathrm{A-U}\choose\mathrm{U-A}}$,
$\Delta H{\mathrm{A-U}\choose\mathrm{C-G}}$ and $\Delta H{\mathrm{U-A}\choose\mathrm{G-C}}$,
$\Delta H{\mathrm{U-A}\choose\mathrm{C-G}}$ and $\Delta H{\mathrm{A-U}\choose\mathrm{G-C}}$,
and similarly for the corresponding entropies.

Based on the experimental data for stacking of nucleic acids in single-stranded RNA (see Table 8.1 in 
ref 3 and Table 1 in ref 4), we make additional assumptions that 
$\Delta H{\mathrm{C}\choose\mathrm{C}}=\Delta H{\mathrm{U}\choose\mathrm{C}}$, 
$\Delta H{\mathrm{A}\choose\mathrm{A}}=\Delta H{\mathrm{U}\choose\mathrm{A}}=\Delta H{\mathrm{C}\choose\mathrm{A}}$, 
and similarly for the entropies. This allows us to combine eq 2 for ${\mathrm{U-A}\choose\mathrm{A-U}}$
with the experimentally determined melting temperature $t_m{\mathrm{A}\choose\mathrm{A}}=26\ ^{\circ}$C\cite{NAbook}, 
which under our assumptions equals $t_m{\mathrm{U}\choose\mathrm{A}}$, and to solve for 
$\Delta H{\mathrm{U}\choose\mathrm{A}}$, $\Delta S{\mathrm{U}\choose\mathrm{A}}$ and $\Delta H(\mathrm{A}-\mathrm{U})$. 
Now putting $\Delta H{\mathrm{A}\choose\mathrm{A}}=\Delta H{\mathrm{U}\choose\mathrm{A}}$ and
$\Delta S{\mathrm{A}\choose\mathrm{A}}=\Delta S{\mathrm{U}\choose\mathrm{A}}$ in eq 2 for 
${\mathrm{A-U}\choose\mathrm{A-U}}$, we can compute $\Delta H{\mathrm{U}\choose\mathrm{U}}$ and 
$\Delta S{\mathrm{U}\choose\mathrm{U}}$. Finally, we assume
\begin{eqnarray}
\Delta H{\mathrm{U}\choose\mathrm{C}}=k\Delta H{\mathrm{U}\choose\mathrm{A}} + (1-k)\Delta H{\mathrm{U}\choose\mathrm{U}},\nonumber\\
\Delta S{\mathrm{U}\choose\mathrm{C}}=k\Delta S{\mathrm{U}\choose\mathrm{A}} + (1-k)\Delta S{\mathrm{U}\choose\mathrm{U}},
\label{CU}
\end{eqnarray}
where $k=0.615$ yields $t_m{\mathrm{U}\choose\mathrm{C}}=13\ ^{\circ}$C, which matches the experimental result for 
$t_m{\mathrm{C}\choose\mathrm{C}}$\cite{NAbook}. The remaining stacking parameters follow directly from 
eq~2 without any additional approximations, if we use the computed hydrogen bond enthalpy 
$\Delta H(\mathrm{A}-\mathrm{U})=-1.47$ kcal/mol for an $\mathrm{A}-\mathrm{U}$ base pair and $3/2$ times this value for 
a $\mathrm{G}-\mathrm{C}$ pair. The resulting stacking parameters, which are used in the present simulations, are given in 
Table S2. The relative stabilities of stacks agree with the experimental data\cite{NAbook,SingleStrandData}, 
identifying $t_m{\mathrm{G}\choose\mathrm{G}}$ and $t_m{\mathrm{U}\choose\mathrm{U}}$ as the highest and lowest melting 
temperatures among all stacks.

We simulated stacking of nucleotide dimers, similar to that shown in Figure S1a, using the stacking potential 
$U_{\mathrm {ST}}$ in eq~1 and $U_{\mathrm {ST}}^0=-h+k_{\mathrm B}(T-T_m)s$, where 
$k_{\mathrm B}$ is the Boltzmann constant, $T$ (K) is the absolute temperature, $T_m$ (K) is the melting temperature of 
each stack (from Table S2) and $h$ and $s$ are adjustable parameters. In our simulations, we computed the stability 
$\Delta G(T)$ of stacks at temperature $T$ as
\begin{equation}
\Delta G(T) = -k_{\mathrm B}T\log{N_1} + k_{\mathrm B}T\log{N_2} + \Delta G_0, 
\label{dGsim}
\end{equation}
where $N_1$ is the number of all stacked configurations for which $U_{\mathrm{ST}}<-k_{\mathrm B}T$ and $N_2$ is the 
number of all unstacked configurations. We adjusted $h$ and $s$ individually for all stacks so that the simulation 
results for $\Delta H$ and $\Delta S$, given by $\Delta G(T)=\Delta H-T\Delta S$, matched the corresponding values in 
Table S2. The stability correction $\Delta G_0$ in eq~4 is assumed to be constant for all stacks 
and accounts for potential discrepancies between $\Delta G(T)$ measured in experiments and its definition used in our 
simulations.  

As an example, in Figure S2 we plot $\Delta G(T)$ obtained in simulations of the ${\mathrm{G}\choose\mathrm{A}}$ 
stack with various values of $h$ and $s = 0$. The simulation results in Figure S2 are shown for $\Delta G_0=0$.
The observed melting temperature $T^*$, defined as $\Delta G(T^*)=0$, increases with $h$ and equals the target melting
temperature $T_m{\mathrm{G}\choose\mathrm{A}}$ in Table S2 when $h=5.98$ kcal/mol. For $s=0$ the entropy loss 
$\Delta S$ associated with stack formation, given by the slope of $\Delta G(T)$ over $T$, is smaller than the value of 
$\Delta S{\mathrm{G}\choose\mathrm{A}}$ specified in Table S2. To correct this, we use 
$U_{\mathrm {ST}}^0=-5.98+k_{\mathrm B}(T-T_m)s$, which does not result in changes in the melting temperature but 
allows us to adjust the slope of $\Delta G(T)$ by adjusting the value of $s$. We find that $s=5.30$ yields 
$\Delta S{\mathrm{G}\choose\mathrm{A}}$ in Table S2. 

We carried out the same fitting procedure for all nucleotide dimers. The resulting parameters $U_{\mathrm{ST}}^0$
for $\Delta G_0=0$ and $\Delta G_0=0.5$ kcal/mol are summarized in Table S3. In our simulations, we use the value 
$\Delta G_0=0.5$ kcal/mol which yields the best agreement with experiments (see discussion in the last section below). 
Note that, although some stacks have equivalent thermodynamic parameters in Table S2, they may require somewhat 
different $U_{\mathrm{ST}}^0$ due to their geometrical differences.

Coarse-grained hydrogen bond interactions $U_{\mathrm{HB}}$ are assigned based on the hydrogen bonds 
present in the original NMR structure. In this work, we carried out independent simulations of the pseudoknot and 
hairpin conformations of the hTR pseudoknot domain (PDB codes 2K96 and 1NA2, respectively). In both cases, we 
generated an optimal network of hydrogen bonds by submitting the NMR structure to the WHAT IF server at 
{\ttfamily http://swift.cmbi.ru.nl}. Each of the generated bonds is modeled by a coarse-grained interaction potential,
\begin{eqnarray}
U_{\mathrm {HB}}=2.286\times\left[1+5(r-r_0)^2+1.5(\theta_1-\theta_{10})^2+1.5(\theta_2-\theta_{20})^2\right.\nonumber\\
\left.+0.15(\psi-\psi_{0})^2+0.15(\psi_1-\psi_{10})^2+0.15(\psi_2-\psi_{20})^2\right]^{-1},
\label{UHB}
\end{eqnarray}
where $r$, $\theta_1$, $\theta_2$, $\psi$, $\psi_1$ and $\psi_2$ for various coarse-grained sites are defined in 
Figure S1. In the case of Watson-Crick base pairs, the equilibrium values $r_0$, $\theta_{10}$, $\theta_{20}$, 
$\psi_0$, $\psi_{10}$ and $\psi_{20}$ are adopted from the coarse-grained structure of an ideal A-form RNA 
helix\cite{IdealRNAsite}. For all other bonds, the equilibrium parameters are obtained by coarse-graining the PDB structure
itself. Equation~5 specifies $U_{\mathrm{HB}}$ for a single hydrogen bond and it must be multiplied by a 
factor of 2 or 3 if the same coarse-grained sites are connected by more than one hydrogen bond. The complex geometry of 
$U_{\mathrm{HB}}$ is the minimum necessary to maintain stable double (and triple) helices in our coarse-grained model.

\section {Comparison of two alternative sets of interaction parameters}

In our simulations, stacking parameters $U_{\mathrm {ST}}^0$ are determined based on the definition of $\Delta G(T)$ 
given in eq~4. The corrective constant $\Delta G_0=0.5$ kcal/mol in eq~4 is 
introduced  to improve quantitative agreement between simulation and experimental melting data of the hTR pseudoknot domain. 
If $\Delta G_0$ were omitted from eq 4, this would result in stronger stacking interactions (see 
Table S3). In order to preserve overall melting temperatures, this increase in the magnitude of 
$U_{\mathrm {ST}}^0$ must be compensated for by a decrease in the strength of hydrogen bonds. 
For $\Delta G_0=0$, the best agreement with experiments is achieved when the prefactor in eq~5 is reduced 
to 2.065. 

In Figure S3 we compare melting data for the hairpin (HP) conformation of the hTR pseudoknot domain for the two 
parameter sets, with and without the corrective constant $\Delta G_0$. The melting profile of the Watson-Crick part of the 
double helix, stem 1 in HP, is hardly affected by the choice of $U_{\mathrm {ST}}^0$. However, we observe a large 
discrepancy for the uridine-rich stem 2, whose melting temperature increases from 50 $^{\circ}$C to 75 $^{\circ}$C if 
$\Delta G_0$ is set to 0 in eq~4. In the latter case, the total melting profile of HP shows only one 
peak at 78 $^{\circ}$C, which does not compare well with two experimental peaks at 50 $^{\circ}$C and 79 
$^{\circ}$C\cite{ExpHP}. We conclude that the corrective constant in eq~4 is crucial for obtaining 
quantitative agreement with the experimental data for HP. 

Melting of secondary structure in the pseudoknot (PK) conformation of the hTR pseudoknot domain is compared in 
Figure S4 for $\Delta G_0=0.5$ kcal/mol and $\Delta G_0=0$. In experiments\cite{ExpPK}, the temperature range for 
melting of stems 1 and 2 in PK is 65--95 $^{\circ}$C and 60--80 $^{\circ}$C, respectively. As shown in Figure S4, 
these experimental data are reproduced well in simulations with $\Delta G_0=0.5$ kcal/mol (weak stacks in 
Table S3 and  $U_{\mathrm {HB}}^0=2.286$). At the same time, for $\Delta G_0=0$ (strong stacks in 
Table S3 and $U_{\mathrm {HB}}^0=2.065$), melting of both stems occurs in the temperature range 
50--95 $^{\circ}$C. For both parameter sets, the overall melting profile of PK has a sharp peak at 70 $^{\circ}$C, in 
agreement with experiments\cite{ExpPK}.

The distance between two peaks in the melting profile of HP increases with $\Delta G_0$ and exceeds the experimental 
distance when $\Delta G_0>0.5$ kcal/mol (data not shown). We therefore conclude that, for both conformations of the hTR 
pseudoknot domain, the parameter model based on $\Delta G_0=0.5$ kcal/mol yields the optimal agreement with experimental 
thermodynamic data.

\section {Crowder-RNA interactions}

Crowder-RNA interactions are modeled by a generalized Lennard-Jones potential,
\begin{eqnarray}
U_{\mathrm{LJ}}(r)&=&\varepsilon\frac{2R_i}{D_0}\left[\left(\frac {D_0}{r + D_0 - D}\right)^{12} 
- 2\left(\frac {D_0}{r + D_0 - D}\right)^{6} + 1\right],\ r\le D, \nonumber \\
U_{\mathrm{LJ}}(r)&=&0,\ r > D, 
\label{POT1}
\end{eqnarray}
where $r$ is the distance between the particles' centers of mass, $D_0=3.2$ \r{A} is the effective penetration depth, $R_i$
is the radius of an RNA coarse-grained bead (values specified above), $r_{\mathrm C}$ is the radius of a crowder,
and $D=R_i+r_{\mathrm C}$. The ratio $2R_i/D_0$ in eq~6 is used to scale the interaction strength 
$\varepsilon = 1$ kcal/mol in proportion to the surface contact area. 

We use the same formula to model RNA-RNA excluded volume interactions, but take $R_i=1.6$ \r{A} for all RNA
beads. In this case, eq~6 becomes a standard (purely repulsive) Lennard-Jones potential,
\begin{eqnarray}
U_{\mathrm{LJ}}(r)&=&\varepsilon
\left[\left(\frac {D_0}{r}\right)^{12} 
- 2\left(\frac {D_0}{r}\right)^{6} + 1\right],\ r\le D_0, \nonumber \\
U_{\mathrm{LJ}}(r)&=&0,\ r > D_0. 
\label{POT2}
\end{eqnarray}
With adjustment of $R_i$ steric clashes between two stacked bases are avoided.

\section {Simulation details}

The RNA and crowder dynamics are simulated by solving the Langevin equation, which for particle $i$ is 
$m_i\ddot{\mathbf{r}}_i=-\gamma_i\dot{\mathbf{r}}_i+\mathbf{F}_i+\mathbf{R}_i$, where $m_i$ is the 
particle mass, $\gamma_i$ is the drag coefficient, $\mathbf{F}_i$ is the conservative force, and $\mathbf{R}_i$ is the 
Gaussian random force, $\left<\mathbf{R}_i(t)\mathbf{R}_j(t^{\prime})\right>=6k_{\mathrm B}T\gamma_i\delta_{ij}\delta(t-t^{\prime})$. 
The drag coefficient is given by the Stokes formula, $\gamma_i=6\pi \eta R_i$, where $\eta$ is the viscosity of the 
medium and $R_i$ is the particle radius. To enhance conformational sampling, we take $\eta=10^{-5}$Pa$\cdot$s, which equals 
approximately 1\% of the viscosity of water. The masses and radii of RNA coarse-grained beads are specified above. The 
masses of crowders scale with their volume, assuming density equal to that of a typical folded protein such as ubiquitin, 
which has the molecular weight $M_{\mathrm r}=8564$ Da and radius $r_{\mathrm C}=1.2$ nm. The Langevin equation is 
integrated using the leap-frog algorithm with time step $\Delta t=2.5$ fs. The length of a simulation run at each 
temperature is 2.5 $\mu$s.

The number of crowders of a given type is computed from its specified volume fraction and the volume of the simulation box. 
In simulations with large crowders ($r_{\mathrm C}= 10.4$ nm, 5.2 nm and 2.6 nm) we use 60 nm as the side of the cubic 
simulation box. For example, the {\it E. coli} mixture contains the volume fractions $\phi_1=0.11$, $\phi_2=0.11$ and
$\phi_3=0.08$ of crowders with $r_{\mathrm C}= 10.4$ nm, 5.2 nm and 2.6 nm, respectively. For a simulation box with side 
60 nm, this yields 5 crowders with $r_{\mathrm C}= 10.4$ nm, 40 with $r_{\mathrm C}= 5.2$ nm and 234 with 
$r_{\mathrm C}= 2.6$ nm (279 crowders in total). In simulations with small crowders ($r_{\mathrm C}= 1.2$ nm 
and 0.6 nm) the number of crowders in the cubic box with side 60 nm becomes very large. To minimize simulation time 
of these systems, the size and shape of the simulation box is adjusted periodically to accommodate RNA in its current 
conformation. To do so, we first align the walls of the box with the RNA axes of inertia. The new position of the RNA
center of mass and new side lengths $L_x$, $L_y$, $L_z$ are computed so that there is at least a 6 nm distance 
between each RNA bead and all six walls. When, as a result of diffusion, the distance between an RNA bead and a wall 
becomes less than 2.4 nm, the box is adjusted again. The size and shape of the simulation box are therefore directly 
coupled to the RNA configuration. For instance, in simulations at 0 $^{\circ}$C, the RNA remains folded and 
$L_x$, $L_y$, $L_z$ fluctuate around 18 nm, 15 nm and 14 nm, respectively (assuming $L_x>L_y>L_z$). 
In simulations at 120 $^{\circ}$C, the RNA is unfolded and the average lengths are $L_x=20$ nm, $L_y=16.5$ nm and $L_z=15$ nm. 
In simulations at intermediate temperatures, when the RNA folds and unfolds, the size of the box undergoes 
large fluctutions between the high-temperature and low-temperature values. The frequency with which the simulation box is 
adjusted also depends on the temperature through RNA diffusion. The box is adjusted approximately every 1200000 steps at 
0 $^{\circ}$C and every 700000 steps at 120 $^{\circ}$C. When the box is adjusted, the number of crowders 
changes with the new box volume in order to keep the volume fractions $\phi$ constant.

%%\bibliography {JACS_sources}

\newpage

\begin{table}
\renewcommand{\baselinestretch}{2}
\caption{Thermodynamic parameters of double-stranded stacks from ref 2.
In the first column, the $5'$ to $3'$ direction is shown by an arrow.}
\bigskip
\bigskip
\begin{tabular}{c|c|c}
\hline\hline
$\uparrow\hspace{-1mm}{b-c\atop a-d}\hspace{-1mm}\downarrow$ &  $\Delta H$, kcal$\,$mol$^{-1}$ 
& $\Delta S$, cal$\,$mol$^{-1}$K$^{-1}$ \\
\hline
$\mathrm{A-U}\atop\mathrm{A-U}$ & $-6.82$ & $-19.0$ \\
$\mathrm{U-A}\atop\mathrm{A-U}$ & $-9.38$ & $-26.7$ \\
$\mathrm{A-U}\atop\mathrm{U-A}$ & $-7.69$ & $-20.5$ \\
$\mathrm{U-A}\atop\mathrm{C-G}$ & $-10.48$ & $-27.1$ \\
$\mathrm{A-U}\atop\mathrm{C-G}$ & $-10.44$ & $-26.9$ \\
$\mathrm{U-A}\atop\mathrm{G-C}$ & $-11.40$ & $-29.5$ \\
$\mathrm{A-U}\atop\mathrm{G-C}$ & $-12.44$ & $-32.5$ \\
$\mathrm{G-C}\atop\mathrm{C-G}$ & $-10.64$ & $-26.7$ \\
$\mathrm{G-C}\atop\mathrm{G-C}$ & $-13.39$ & $-32.7$ \\
$\mathrm{C-G}\atop\mathrm{G-C}$ & $-14.88$ & $-36.9$ \\
\hline\hline 
\end{tabular}
\label{table1}
\end{table}

\begin{table}
\renewcommand{\baselinestretch}{2}
\caption{Thermodynamic parameters of single-stranded stacks, derived in this work. The matching enthalpies of 
hydrogen bond formation in Watson-Crick base pairs are given in last two rows. The melting temperatures of stacks
are indicated in $^{\circ}$C, $t_m(^{\circ}\mathrm{C})=T_m(\mathrm{K})-273.15$. In the first column, the $5'$ to $3'$ 
direction is shown by an arrow. }
\bigskip
\bigskip
\begin{tabular}{c|c|c|c}
\hline\hline
$\uparrow\hspace{-1mm}{b\atop a}$ &  $\Delta H$, kcal$\,$mol$^{-1}$ & $\Delta S$, cal$\,$mol$^{-1}$K$^{-1}$ & $t_m$,$^{\circ}$C \\ 
\hline
$\mathrm{U}\atop\mathrm{U}$ & $-1.81$ & $-7.2$ & $-21$ \\
$\mathrm{C}\atop\mathrm{C}$ & $-2.87$ & $-10.0$ & 13 \\
$\mathrm{C}\atop\mathrm{U}$; $\mathrm{U}\atop\mathrm{C}$  & $-2.87$ & $-10.0$ & 13 \\ 
$\mathrm{A}\atop\mathrm{A}$ & $-3.53$ & $-11.8$ & 26\\
$\mathrm{A}\atop\mathrm{U}$; $\mathrm{U}\atop\mathrm{A}$ & $-3.53$ & $-11.8$ & 26 \\
$\mathrm{A}\atop\mathrm{C}$; $\mathrm{C}\atop\mathrm{A}$ & $-3.53$ & $-11.8$ & 26\\ 
$\mathrm{G}\atop\mathrm{C}$ & $-4.21$ & $-13.3$ & 42\\
$\mathrm{G}\atop\mathrm{U}$; $\mathrm{U}\atop\mathrm{G}$ & $-5.55$ & $-16.4$ & 65 \\
$\mathrm{C}\atop\mathrm{G}$ & $-6.33$ & $-18.4$ & 70\\ 
$\mathrm{G}\atop\mathrm{A}$; $\mathrm{A}\atop\mathrm{G}$ & $-6.75$ & $-19.8$ & 68 \\
$\mathrm{G}\atop\mathrm{G}$ & $-8.31$ & $-22.7$ & 93 \\
\hline
\multicolumn{4}{c}{$\Delta H(\mathrm{A}-\mathrm{U})=-1.47$ kcal/mol }\\
\multicolumn{4}{c}{$\Delta H(\mathrm{G}-\mathrm{C})=-2.21$ kcal/mol }\\
\hline\hline 
\end{tabular}
\label{table2}
\end{table}

\begin{table}
\renewcommand{\baselinestretch}{2}
\caption{Temperature-dependent stacking parameters $U_{\mathrm {ST}}^0$ used in eq~1. The two sets of 
$h$ correspond to two different values of the additive constant $\Delta G_0$ in eq~4: 0.5 kcal/mol 
and 0 (numbers in brackets). The melting temperatures $T_m$ of individual stacks are given in Table S2. The values 
of $s$ conform to $k_{\mathrm B}(T-T_m)$ evaluated in kcal$\,$mol$^{-1}$. In the first column, the $5'$ to $3'$ direction 
is shown by an arrow.}
\bigskip
\bigskip
\begin{tabular}{c|c|c}
\hline\hline
& \multicolumn{2}{|c}{$U_{\mathrm {ST}}^0=-h+k_{\mathrm B}(T-T_m)s$}\\
\hline
$\uparrow\hspace{-1mm}{b\atop a}$ &  $h$, kcal$\,$mol$^{-1}$ & $s$ \\ 
\hline
$\mathrm{U}\atop\mathrm{U}$ & 3.52 (4.27) & $-3.56$\\
$\mathrm{C}\atop\mathrm{C}$ & 4.16 (4.87) & $-1.57$ \\
$\mathrm{C}\atop\mathrm{U}$; $\mathrm{U}\atop\mathrm{C}$ & 4.14 (4.88); 4.14 (4.87) & $-1.57$; $-1.57$\\ 
$\mathrm{A}\atop\mathrm{A}$ & 4.49 (5.19) & $-0.32$\\
$\mathrm{A}\atop\mathrm{U}$; $\mathrm{U}\atop\mathrm{A}$ & 4.43 (5.16); 4.45 (5.15) & $-0.32$; $-0.32$\\
$\mathrm{A}\atop\mathrm{C}$; $\mathrm{C}\atop\mathrm{A}$ & 4.43 (5.16); 4.45 (5.15) & $-0.32$; $-0.32$\\ 
$\mathrm{G}\atop\mathrm{C}$ & 4.75 (5.48) & 0.77\\ 
$\mathrm{G}\atop\mathrm{U}$; $\mathrm{U}\atop\mathrm{G}$ & 5.17 (5.89); 5.12 (5.84) & 2.92; 2.92\\
$\mathrm{C}\atop\mathrm{G}$ & 5.22 (5.93) & 4.37\\
$\mathrm{G}\atop\mathrm{A}$; $\mathrm{A}\atop\mathrm{G}$ & 5.26 (5.98); 5.22 (5.95) & 5.30; 5.30\\
$\mathrm{G}\atop\mathrm{G}$ & 5.70 (6.42) & 7.35\\
\hline\hline 
\end{tabular}
\label{table3}
\end{table}

\newpage

\begin{figure}
\includegraphics[width=13.5cm,clip]{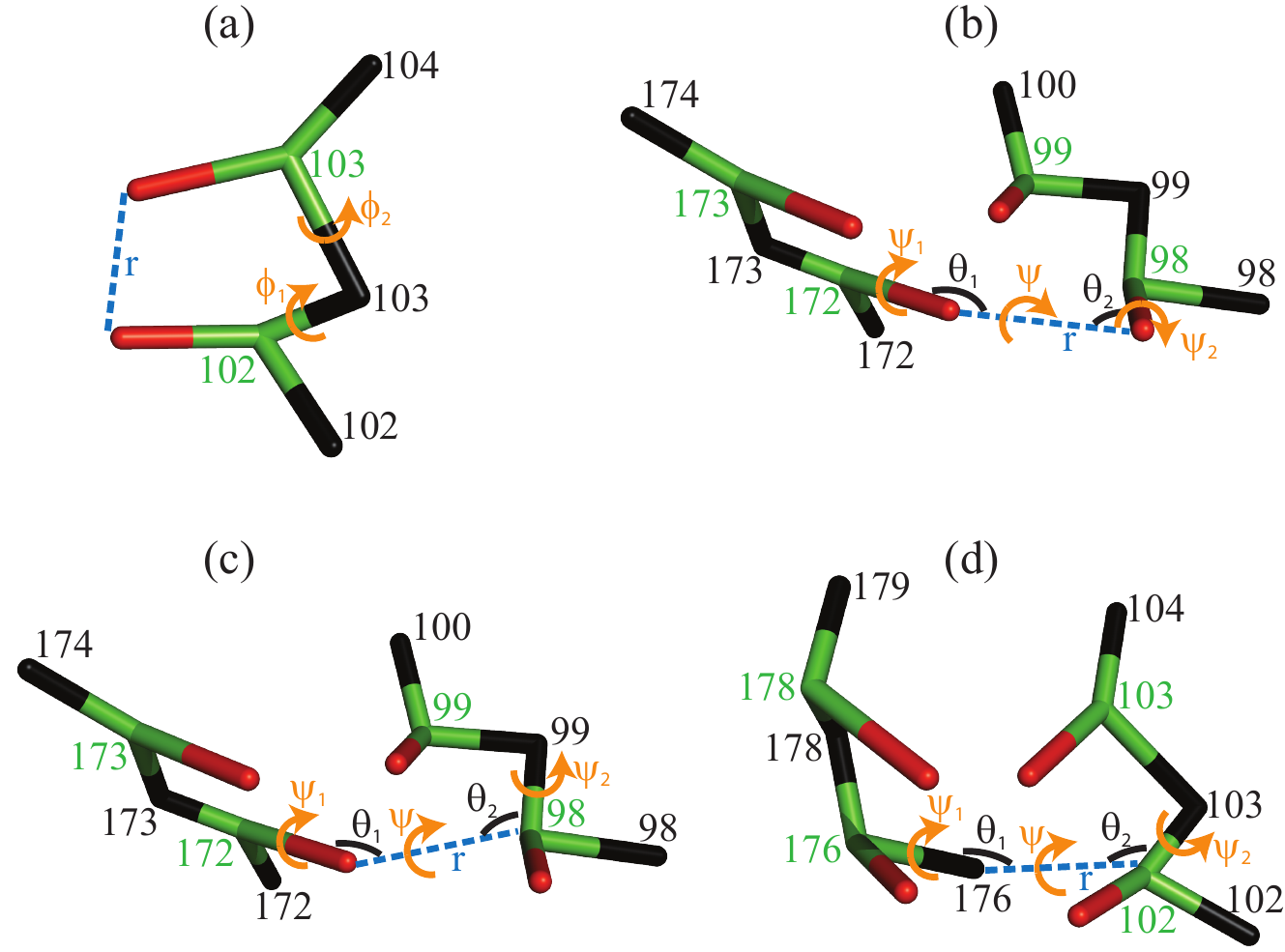}
\renewcommand{\baselinestretch}{2}
\caption{Geometrical parameters for (a) stacking in eq~1 and (c--d) hydrogen bonding in eq~5. 
Sites P, S and B are shown in black, green and red, respectively. Sample conformations are derived from 2K96.pdb. 
The numbers refer to specific nucleotides, and $r$, $\theta$, and $\phi$ or $\psi$ refer to distances (dist.), bond angles 
(ang.) and dihedral angles (dih.) between indicated sites. (a) Stacking: $r$ = dist. (B102, B103), 
$\phi_1$ = dih. (P102, S102, P103, S103), $\phi_2$ = dih. (P104, S103, P103, S102). 
(b) Hydrogen bonding between B and B: $r$ = dist. (B98, B172), $\theta_1$ = ang. (S172, B172, B98), 
$\theta_2$ = ang. (S98, B98, B172), $\psi$ = dih. (S98, B98, B172, S172), $\psi_1$ = dih. (B98, B172, S172, P173), 
$\psi_2$ = dih. (B172, B98, S98, P99). (c) Hydrogen bonding between B and S: 
$r$ = dist. (S98, B172), $\theta_1$ = ang. (S172, B172, S98), $\theta_2$ = ang. (P99, S98, B172), 
$\psi$ = dih. (P99, S98, B172, S172), $\psi_1$ = dih. (S98, B172, S172, P173), 
$\psi_2$ = dih. (B172, S98, P99, S99). (d) Hydrogen bonding between P and S: $r$ = dist. (S102, P176), 
$\theta_1$ = ang. (S176, P176, S102), $\theta_2$ = ang. (P103, S102, P176), $\psi$ = dih. (P103, S102, P176, S176), 
$\psi_1$ = dih. (S102, P176, S176, P178), $\psi_2$ = dih. (P176, S102, P103, S103).}
\label{HBONDS}
\end{figure}

\begin{figure}
\includegraphics[width=8.0cm,clip]{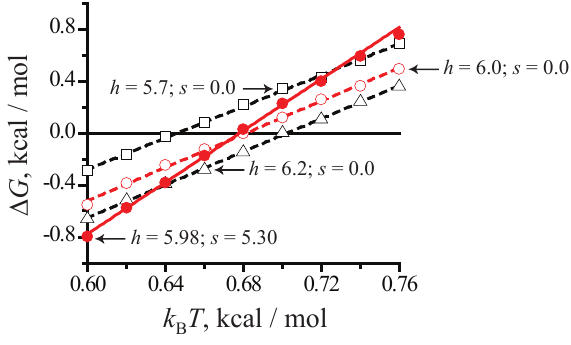}
\renewcommand{\baselinestretch}{2}
\caption{ Stability $\Delta G(T)$ of stack ${\mathrm{G}\choose\mathrm{A}}$, computed from eq~4 with
$\Delta G_0=0$, in simulations with the stacking potential given by eq~1 and 
$U_{\mathrm {ST}}^0=-h+k_{\mathrm B}(T-T_m)s$. $T_m$ indicates the melting temperature for ${\mathrm{G}\choose\mathrm{A}}$ 
from Table S2, $k_{\mathrm B}T_m=0.68$ kcal/mol. Open and closed symbols show simulation results for various $h$ 
and $s$. Red solid line indicates the target stability $\Delta G(T)=\Delta H-T\Delta S$ for 
${\mathrm{G}\choose\mathrm{A}}$, where $\Delta H$ and $\Delta S$ are given in Table S2. Same stability line is 
obtained in simulation with $h=5.98$ kcal/mol and $s=5.30$ (closed symbols).}
\label{UST0}
\end{figure}

\begin{figure}
\includegraphics[width=13.5cm,clip]{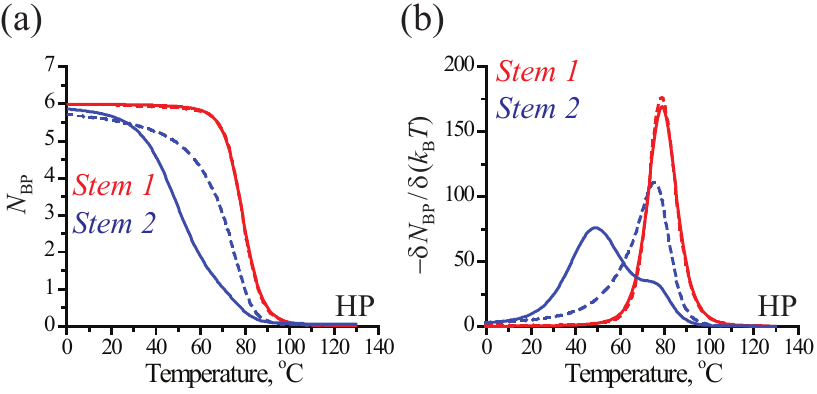}
\renewcommand{\baselinestretch}{2}
\caption{ (a) Number of intact base pairs $N_{\mathrm{BP}}$ vs. temperature in two elements of secondary structure in HP. 
(b) The melting profiles show the rate of change of the number of intact base pairs with temperature. The data are 
obtained for different values of $\Delta G_0$ in eq~4: 0.5 kcal/mol (solid) and 0 (dashed).}
\label{HPDU}
\end{figure}

\begin{figure}
\includegraphics[width=13.5cm,clip]{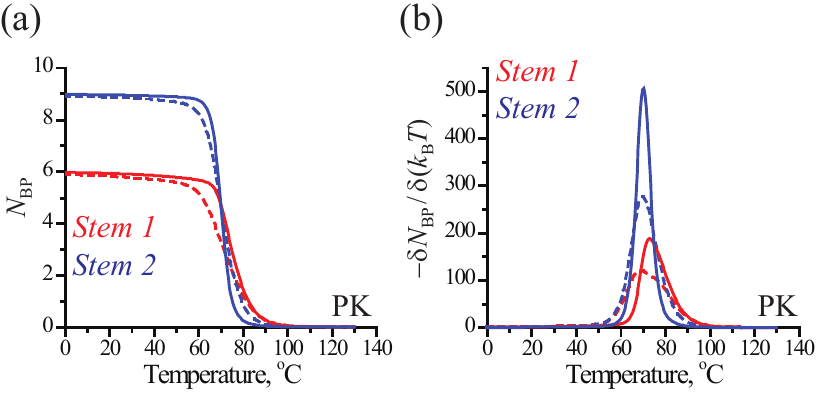}
\renewcommand{\baselinestretch}{2}
\caption{Same as in Figure S3, but for the PK conformation.}
\label{PKDU}
\end{figure}

\begin{figure}
\includegraphics[width=13.5cm,clip]{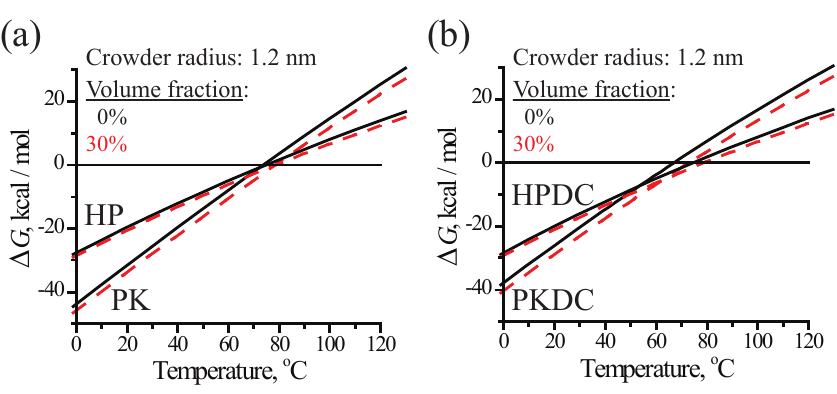}
\renewcommand{\baselinestretch}{2}
\caption{Temperature dependence of the stabilities of the hairpin and pseudoknot in the absence of crowders (solid) and in 
a monodisperse suspension of crowders with $\phi=0.3$ and $r_{\mathrm{C}}=1.2$ nm (dashed). (a) $\Delta$U177 sequence. 
(b) $\Delta$U177 sequence with DKC mutations.}
\label{dG}
\end{figure}

\begin{figure}
\includegraphics[width=11.0cm,clip]{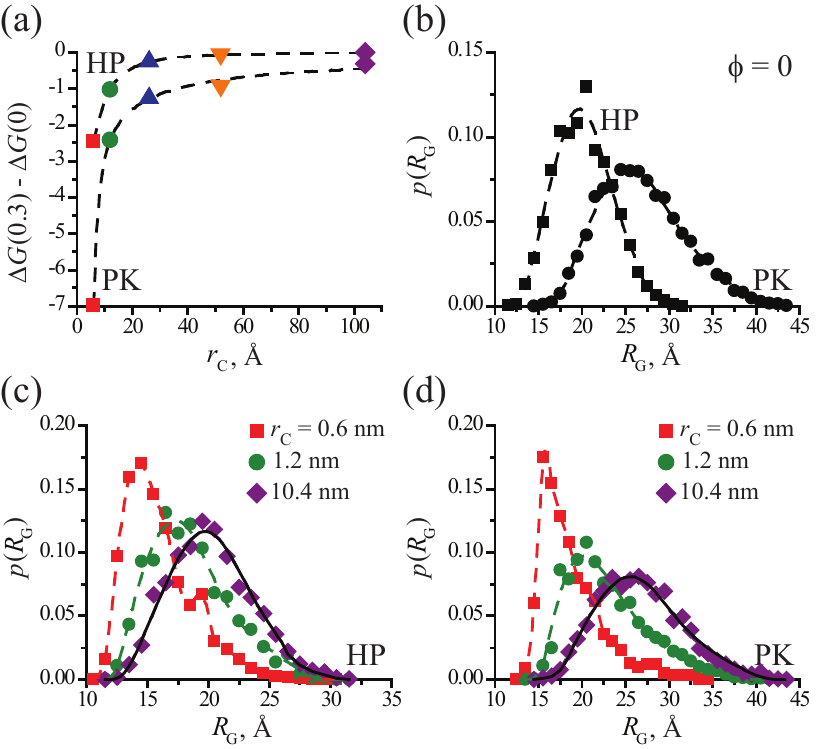}
\renewcommand{\baselinestretch}{2}
\caption{(a) Changes in stability (kcal/mol) of HP and PK at 37 $^{\circ}$C due to crowders, as a function of the crowder 
radius $r_{\mathrm{C}}$. (b) Probability distributions $p(R_{\mathrm{G}})$ of the radius of gyration of the unfolded HP 
and PK structures at 37 $^{\circ}$C in the absence of crowders ($\phi=0$). $R_{\mathrm{G}}$ of HP is defined as the
radius of gyration of strand G93--C121, which is the only fragment that is structured in HP (same definition is used in 
the paper). $R_{\mathrm{G}}$ of PK is computed for the entire length of RNA. The unfolded state cannot be sampled directly 
in simulations at 37 $^{\circ}$C and the shown $p(R_{\mathrm{G}})$ are obtained by statistical reweighting of the 
high-temperature data. For PK, $p(R_{\mathrm{G}})$ is noticeably asymmetric, showing an extended tail on the 
side of large $R_{\mathrm{G}}$. This is in part due to residual stacking in the unfolded state, whose effects become more 
pronounced for longer strands. A larger mean of $p(R_{\mathrm{G}})$ and its asymmetry explain why the 
PK structure is more strongly influenced by crowders than HP. (c)--(d) $p(R_{\mathrm{G}})$ in crowder suspensions with 
$\phi=0.3$ and different $r_{\mathrm{C}}$. Symbols match those in panel (a) and black solid curves show 
$p(R_{\mathrm{G}})$ at $\phi=0$. In all four panels, dashed curves are drawn through data to facilitate comparison.}
\label{RG}
\end{figure}

\begin{figure}
\includegraphics[width=7.0cm,clip]{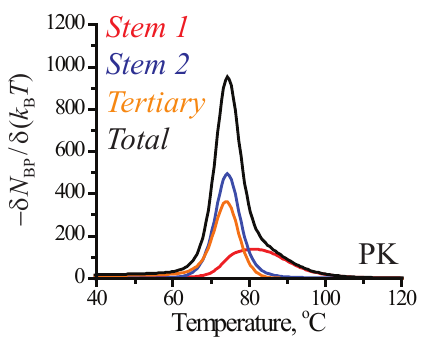}
\renewcommand{\baselinestretch}{2}
\caption{Same as in Figure 2a of the main text, but in a monodisperse suspension of crowders with $\phi=0.3$ and 
$r_{\mathrm{C}}=1.2$ nm.}
\label{PKCRW}
\end{figure}

\begin{figure}
\includegraphics[width=7.0cm,clip]{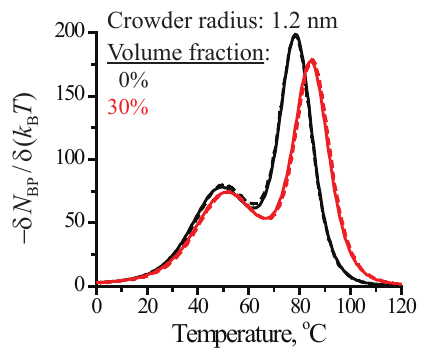}
\renewcommand{\baselinestretch}{2}
\caption{Melting profiles of the hairpin in the absence of crowders (black) and in a monodisperse suspension of crowders 
with $\phi=0.3$ and $r_{\mathrm{C}}=1.2$ nm. Dashed and solid curves are for the $\Delta$U177 sequence with and 
without additional DKC mutations. }
\label{HPDC}
\end{figure}

\end {document}